\documentstyle[12pt,aasms4,psfig]{article}
\newcommand{\apgt}{\ {\raise-.5ex\hbox{$\buildrel>\over\sim$}}\ }
\newcommand{\aplt}{\ {\raise-.5ex\hbox{$\buildrel<\over\sim$}}\ }

\def\lta{{\>\rlap{\raise2pt\hbox{$<$}}\lower3pt\hbox{$\sim$}\>}}
\def\gta{{\>\rlap{\raise2pt\hbox{$>$}}\lower3pt\hbox{$\sim$}\>}}

\lefthead{Livio and Stiavelli}
\righthead{Does the Fine Structure Constant Really Vary in Time?}

\begin{document}

\title{Does the Fine Structure Constant Really Vary in Time?}
\author{Mario Livio and Massimo Stiavelli\footnote{On leave from the
Scuola Normale Superiore, Piazza dei Cavalieri 7, I56126 Pisa,
Italy}$^,$\footnote{On assignment from the Space Science Dept. of the
European Space Agency}\\ 
Space Telescope Science Institute\\ 
3700 San Martin Drive\\ Baltimore, MD 21218}

\begin{abstract}
We discuss how laboratory experiments can be used to place constraints
on possible variations of the fine structure constant $\alpha$ in the
observationally relevant redshift interval $z \simeq 0 - 5$, within a
rather general theoretical framework. We find a worst case upper limit
for $\Delta \alpha / \alpha$ of $8 \times 10^{-6}$ for $z \leq 5$ and
$\Delta \alpha / \alpha$ of $0.9 \times 10^{-6}$ for $z \leq 1.6$. The
derived limits are at variance with the recent findings by Webb
et~al.\ (1998), who claim an observed variation of $\Delta
\alpha/\alpha = -2.6 \pm 0.4 \times 10^{-5}$ at $1<z<1.6$.
\end{abstract}

{\it subject headings}: cosmology: theory - physical data and theory:
relativity, nuclear reactions, nucleosynthesis, abundances

\section{Introduction}

In an interesting paper, Webb et~al.\ (1998) describe an observational
method for investigating possible time and space variations in the fine
structure constant $\alpha$, using quasar absorption systems (see also
Drinkwater et~al.\ 1998).  Webb et~al.\ present intriguing evidence
suggesting that $\alpha$ was smaller in the past (for redshifts $z >
1$).  The claimed fractional change is (for $z > 1$), $\Delta
\alpha/\alpha = -2.6 \pm 0.4 \times 10^{-5}$.  

Since a varying value of $\alpha$ can, in principle at least, have
significant consequences for cosmology (e.g.\ concerning
recombination), we examine in the present letter possible variations
in $\alpha$ in a broader context, although still under a particular
set of assumptions.  We also show that in the context of the theory
presented in this paper the result obtained by Webb et~al.\ appears to
be in conflict with results of tests of the equivalence principle.

\section{By How Much Can $\alpha$ Change?}

The experimental constraints on variations of $\alpha$ were explored
extensively by Bekenstein (1982) by constructing a dynamical theory
relating variations of $\alpha$ to the electromagnetic fraction of the
mass density in the universe.  Bekenstein's (1982) framework for
$\alpha$ variability was based on {\it very general assumptions}:
covariance, gauge invariance, causality, time reversal of
electromagnetism, and that gravitation is described by a metric of
spacetime which satisfies Einstein's equations.  He obtained the
following equation for the temporal variation of $\alpha$ (adopting a
Robertson-Walker metric; we correct here a misprint in the original
paper)
\begin{equation}
\label{eq:first}
(a^3 \dot \epsilon/\epsilon{\dot)} = -a^3\zeta (l^2/\hbar c)\rho_{m}
c^4~~~.
\end{equation}
Here $\epsilon = (\alpha/\alpha_{\rm today})^{1/2}$, $l$ is a length
scale of the theory, $\rho_m$ is the total rest mass density of
matter, $a$ is the expansion scale factor, and $\zeta = \langle
m_{n,em} \rangle / m_p$ is a dimensionless parameter which measures
the fraction of mass in the form of Coulomb energy of an average
nucleon, $ \langle m_{n,em} \rangle$, compared to the free proton
mass, $m_p$.  In order to be able to integrate eq.~(1), Bekenstein
(1982) assumed that $\zeta$ (which is affected by transformations of
hydrogen into heavier nuclei in stars) is constant.  He was able to
show that this assumption is reasonable up to redshifts $z \aplt 1$,
by the following consideration.  The proton conversion rate (via the
reaction 4H$^1 \to$~He$^4$) can be estimated from the mean luminosity
density (e.g.\ Davis, Geller \& Huchra 1978) to be $\sim 8 \times
10^{-28}$~cm$^{-3}$~s$^{-1}$.  Since a minimal nucleon number density
in the universe can be determined from nucleosynthesis to be (e.g.\
Olive, Schramm \& Steigman 1981) $\sim 10^{-7}$~cm$^{-3}$, the
timescale for proton conversion (and thereby for a change in $\zeta$)
is $\tau_{\rm conv} \apgt 10^{20}s$.

The timescale for changes in $\rho_m$ or $a$ is $\sim H_o^{-1}
\sim 10^{17}s$.  Thus, for as far back as the luminosity of galaxies
does not change significantly (up to $z \aplt 1$), the assumption of a
constant $\zeta$ is a reasonable one.

However, it is interesting to push this analysis to higher redshifts
(e.g. the main change in $\alpha$ claimed by Webb et~al.\ (1998)
occurs at $z \apgt 1$) by extending Bekenstein's (1982) analysis with
a determination of the behavior of $\zeta$ at higher redshifts.  We
achieve this as follows.  As explained above, the rate of change of
the Coulomb contribution to the baryon mass depends on the rate of
conversion of hydrogen into helium. Following Bekenstein we find:
\begin{equation}
\label{eq:zeta}
\zeta \simeq 1.2 \times 10 ^{-2} (X + \frac{4}{3} Y)~~~, 
\end{equation}
where $X$ and $Y$ are the mass fractions in hydrogen and helium,
respectively. In order to determine the conversion rate we adopted
three possible star formation rate (SFR) histories for the universe:
(i)~one that follows the Madau et~al.\ (1996) curve, (ii)~one that
rises from the present up to a peak at $z \sim 1$ and then is flat at
the peak value up to $z \sim 5$ (as is perhaps suggested by the COBE
Diffuse Infrared Background experiment; Hauser et~al.\ 1998; Calzetti
\& Heckman 1998), and (iii)~a SFR that is constant (at the peak value)
for all redshifts (as a limiting case).  These star formation
histories were used to synthesize a mean stellar population of the
universe as a function of time, using the Bruzual and Charlot (1993)
evolutionary models. The bolometric luminosity from the mean stellar
population was then used to calculate the hydrogen to helium
conversion rate.  In Fig.~1, we show the proton conversion rate as a
function of redshift up to $z = 5$, for the three assumed SFR
histories.  As can be seen from the figure, the timescale for proton
conversion satisfies $\tau_{\rm conv} \apgt 10^{19}s$ up to $z = 5$.
Since $\tau_{\rm conv} \gg H_o^{-1}$, we can still take $\zeta$ to be
constant in integrating eq.~(1). Indeed direct calculation of
eq.~(\ref{eq:zeta}) shows that it varies by less than 1 \% over the
redshift interval that we consider. Thus we obtain:
\begin{equation}
\label{eq:derivative}
\dot\epsilon/\epsilon = - \zeta (l^2 c^3/\hbar) \rho_m (t-t_c)~~~,
\end{equation}
where $t_c$ is an (unknown) integration constant. From
eq.~(\ref{eq:zeta}) with $X=0.74$ and $Y = 0.26$ we find for $\zeta$
the value $1.3 \times 10^{-2}$ (see also Bekenstein 1982).  We will
consider in the following the case of a low $\Omega$ ($\Omega \simeq
0.2$, see below) Universe. The results however would not change
significantly in a $\Omega=1$ Universe.  Eq.~(\ref{eq:derivative}) can
be integrated in time by taking into account that for $z <<
4(\Omega^{-1}-1)$ one has $\rho_m \simeq \Omega_{mo} \rho_{c}
(t_o/t)^3$, where $\Omega_{mo}$ is the baryon fraction at the present
time $t_0$, and $\rho_c$ is the critical density $\rho_c \equiv 3
H_0^2/(8\pi G)$, and that $|\epsilon-1| << 1$. We find:
\begin{equation}
\label{eq:generalresult}
\epsilon(z) - 1 \simeq - \frac{3 \zeta}{8 \pi} \left(\frac{l}{L_p}\right)^2
\Omega_{mo} H_o^2 t_o^2 \left(1+\frac{t_c t_0}{2 t^2(z)}-\frac{t_o}{t(z)}-
\frac{t_c}{2 t_o} \right)~~~,
\end{equation}
where $L_p = (G \hbar/c^3)^{1/2}$ is the Planck length. We have
verified that the above analytical approximation is applicable to the
redshift interval of interest. As an example, in Fig.~2 we plot for
the case $t_c = t_0$ the difference between the expression of eq.~(4)
and the result of a direct numerical integration of eq.~(3).
Eq.~(\ref{eq:generalresult}) has two quite different regimes depending
on the value of $t_c/t_o$. Let us consider first the case $|t_c| \lta
t_o$ which is perhaps the most physically founded since there is no a
priori reason for having $|t_c| >> t_o$.  We find:
\begin{equation}
\label{eq:notc}
|\epsilon(z\leq 5) - 1| \lta 6.2 \times 10^{-3} \Omega_{mo} \left(\frac{l}{L_p}
\right)^2~~~.
\end{equation}
In order to obtain an absolute upper limit on the variability, we will
assume that {\it all} the matter density is nucleonic, by taking
$\Omega = \Omega_{mo} \simeq 0.2$ (e.g.\ Garnavich et~al.\ 1998; Perlmutter
et~al.\ 1998). We can also use the results of the
E\"otv\"os-Dicke-Braginsky experiments (E\"otv\"os, Parker \& Fekete
1922; Roll, Krotkov \& Dicke 1964; Braginsky \& Pamov 1972), designed
to test the equivalence principle, to infer $l/L_p \aplt 10^{-3}$.
Thus, our final upper limit is:
\begin{equation}
\label{eq:redshiftfive}
|\epsilon(z\leq 5) - 1| \lta 1.2 \times 10^{-9}~~~.
\end{equation}

If we assume instead that $|t_c| >> t_o$ we can set an upper limit to
$|t_c|$ by considering constraints on $|\dot\epsilon/\epsilon|$. From
Shlyakhter's (1976) analysis of the Oklo natural reactor we have
$|\dot\epsilon/\epsilon| < 0.5 \times 10^{-7} H_o$. More recently,
Damour and Dyson (1996) have derived the less strong but more robust
limit of $|\dot\epsilon/\epsilon| <3.4  \times 10^{-7} H_o$, which
we will adopt in the following. When used in
combination with eq.~(\ref{eq:derivative}) and with $\Omega_{mo} =
0.2$ this gives:
\begin{equation}
\label{eq:tclimits}
| (l/L_p)^2 t_c H_0 | < 6.6 \times 10^{-3}~~~.
\end{equation}
The term involving $t_c$ in Eq.~(\ref{eq:generalresult}) becomes
dominant and we obtain:
\begin{equation}
\label{eq:againfive}
|\epsilon(z\leq 5) - 1| \lta 8 \times 10^{-6}~~~.
\end{equation}

The above limits become stronger as the redshift decreases. For
example, at the redshift of interest for the Webb et al. (1998) result
(z=1.6) one would find instead $|\epsilon(z=1.6)-1| \lta 1.6 \times
10^{-10}$ if $|t_c|/t_o \lta 1$ or $|\epsilon(z=1.6)-1| \lta 9 \times
10^{-7}$ otherwise. Both limits, regardless of the value of $t_c$, are
at variance with the Webb et~al.\ (1998) result.

\section{Discussion}

The analysis of experimental constraints for the variation of the fine
structure constant within the framework of rather general dynamical
theories provides upper limits to any change in $\alpha$ which are at
variance with the Webb et.~al\ (1998) detection. We have verified that
varying the cosmological parameters (i.e. the values of $\Omega$, $H_0$,
primordial helium fraction) does not affect our
constraints significantly.

We should note that variations in $\alpha$ in the context of the theory 
discussed in the present paper result in no modifications to the Planckian
spectrum of black body radiation. In fact one could think of varying $\alpha$ 
as a varying permittivity of the vacuum. This would simply be equivalent to 
changing the expansion factor $a(t)$ to a different function, which cannot
modify the Planck spectrum.

Variations of the fine structure constant $\alpha$ could be driven
also by a different dynamics. In theories with extra dimensions (e.g.,
Kaluza-Klein or superstrings theories) variations in the observed
value of $\alpha$ can be induced by variations in the length scale of
the compact dimensions (see, e.g., Marciano 1984, Barrow 1987). The
effects of extra dimensions were discussed in detail by Barrow (1987)
who found upper limits to $|\dot \alpha / \alpha|$ of $\lta 10^{-8}
H_0 $ for both Kaluza-Klein and superstring theories. This upper limit
is also much smaller than the claimed detection by Webb et.~al\
(1998). However, note that in these theories the variations of all
physical constants are inter-related.

Finally, we would like to mention that the calculations leading to
eq.~(\ref{eq:generalresult}) would have to be modified if a
significant component of the dark matter in the universe were to decay
electromagnetically at redshift $z \apgt 1$ (e.g.\ Sciama 1982,
1997).  We hope that the present letter will inspire more attempts to
determine possible variations of $\alpha$ observationally.

\acknowledgments ML acknowledges support from NASA Grant NAG5-6857.
We thank Yacov Bekenstein for very useful discussions and an anonymous
referee for a careful reading of the manuscript.

\newpage
\begin{figure}
%\centerline{\psfig{figure=luminosity.ps}}
\label{figure1}
\caption{We plot the proton conversion rate as a function of redshift
for a constant SF model (solid line), the Madau et.~al\ cosmic SFR
history (dotted line) and a SFR history rising up to $z=1$ and then staying
constant at the peak value (dashed line). }
\end{figure}

\begin{figure}
%\centerline{\psfig{figure=analytical.ps}}
\label{figure2}
\caption{Difference between the analytical approximation and the
numerically integrated value of $\epsilon$. For redshift $z \leq10$ the
difference between the two is negligible.}
\end{figure}

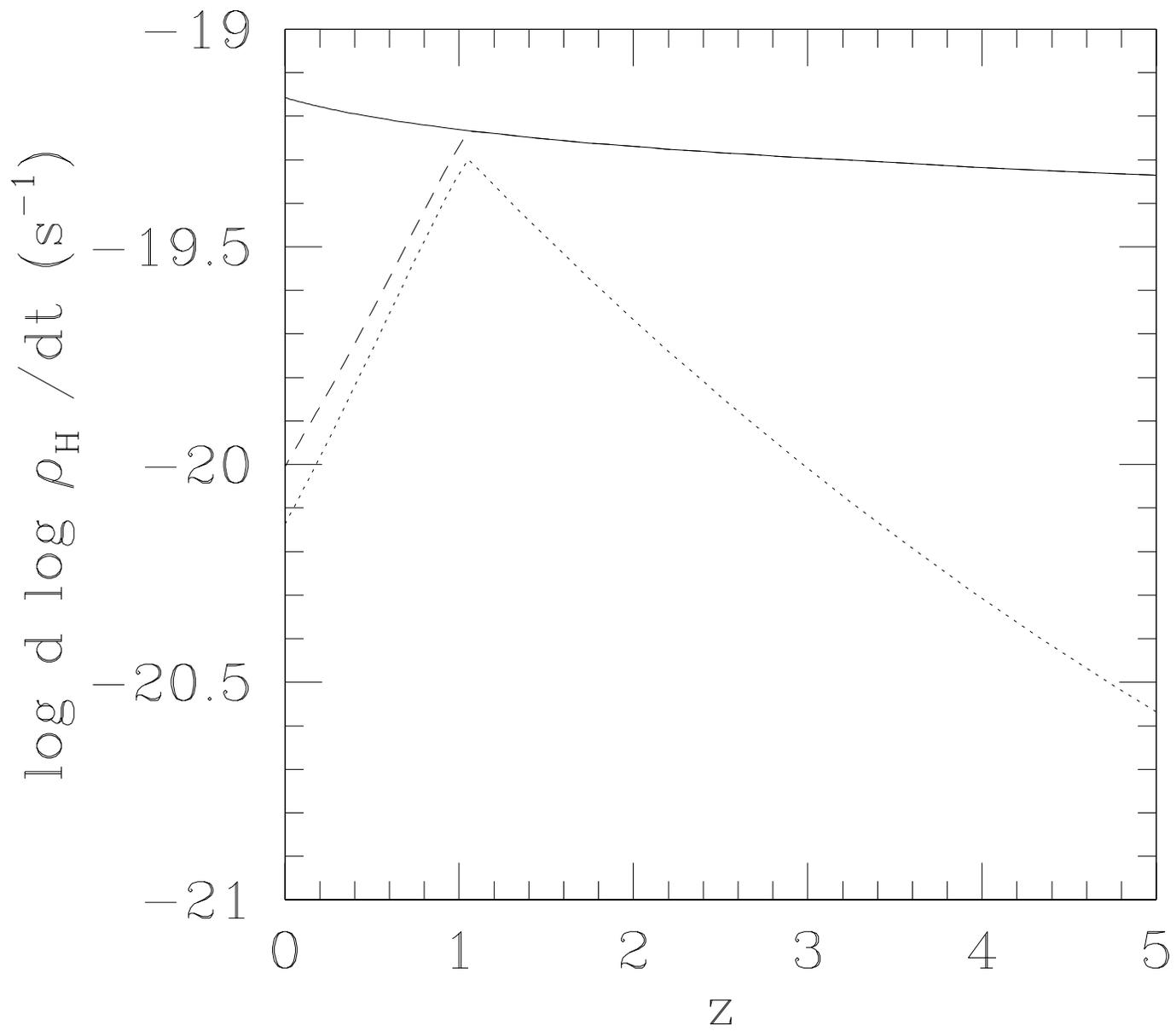
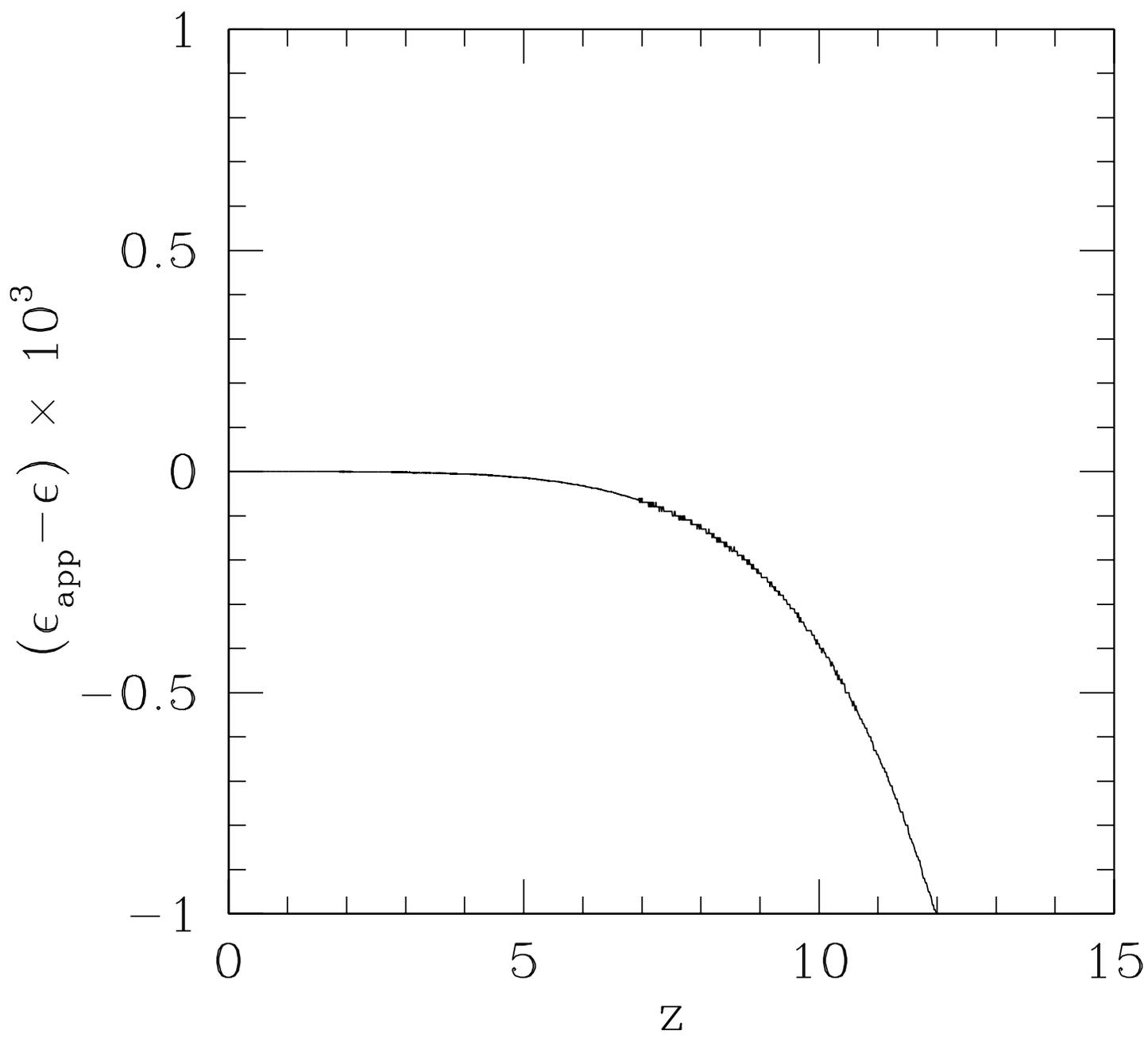
\end{document}